\documentclass[twocolumn,showpacs,preprintnumbers,amsmath,amssymb,epsfig]{revtex4}
\usepackage{epsfig}
\begin{document}

\def\omunit{{(km s$^{-1}$)/kpc}}
\def\gtrsim{ \lower .75ex \hbox{$\sim$} \llap{\raise .27ex \hbox{$>$}} }
\def\lesssim{ \lower .75ex \hbox{$\sim$} \llap{\raise .27ex \hbox{$<$}} }

\def\sun{\odot}
\def\A{{\sf A}}
\def\B{{\sf B}}
\def\Ms{\mathrm{M}_\sun}
\def\vk{v_\mathrm{kick}}
\def\ee{{\it e}}
\newcommand\arcsec{\mbox{$^{\prime\prime}$}}%

\def\figurestyle#1{\sf \small \baselineskip 14pt #1}


 \title{The Origin of the Binary Pulsar J0737-3039B
 }
 \author{Tsvi Piran$^{1,2}$ \& Nir J. Shaviv$^{1}$}
\email{shaviv@phys.huji.ac.il, piran@phys.huji.ac.il}
 \affiliation{
1. Racah Institute of Physics, Hebrew University, Jerusalem 91904,
Israel \\
2. Theoretical Astrophysics 130-33, Caltech, Pasadena, CA 91125,
USA}
\begin{abstract}

Evolutionary scenarios suggest that the progenitor of the new
binary pulsar J0737-3039B \cite{ref1,ref2} was a He-star with $M >
2.1-2.3~\Ms$ \cite{ref3,ref4}. We show that this case implies that
the binary must have a large ($>120$~km/s) center of mass
velocity. However, the location, $\sim 50$~pc from the Galactic
plane, suggests that the system has, at high likelihood, a
significantly smaller center of mass velocity and a progenitor
more massive than 2.1~$\Ms$ is ruled out (at 97\% c.l.). A
progenitor mass around 1.45~$\Ms$, involving a new previously
unseen gravitational collapse, is kinematically favored. The low
mass progenitor is consistent with the recent scintillations based
velocity measurement of 66$\pm 15$~km/s \cite{ref12new} (and which
also rules out the high mass solution at 99\% c.l.)  and
inconsistent with the higher earlier estimates of 141$\pm
8.5$~km/s \cite{ref11new}. Direct proper motion measurements, that
should be available within a year or so, should better help to
distinguish between the two scenarios.

\end{abstract}

%






\maketitle


The remarkable binary system J0737-3039 \cite{ref1,ref2} is
composed of two pulsars denoted \A\ and \B. We show here that the
orbital parameters of this system and its location close to the
galactic plane pose strong limits on the origin of this binary
system and on the progenitor's mass of the younger pulsar \B.

The separation, $R$, (i.e., the sum of the semi-major axes) of the
pulsars today is $8.8 \times 10^{10}$cm and the eccentricity,
$\ee$, is 0.087779. Both $R$, and $\ee$ decrease with time due to
gravitational radiation emission and the system will merge in
approximately $85$~Myr from now. The periods $P_{A,B}$ and their
time derivatives provide upper limits for the life times of the
pulsars: $t_A \approx 210$~Myr and $t_B \approx 50$~Myr.
Integration backwards in time \cite{Peters} yields the system's
parameters at birth, $50$~Myr ago when \B\ was born: $R \approx
10^{11}$cm and $e \approx 0.11$, both only slightly larger than
the present values. The values $210$~Myr ago were not that
different: $R=1.2 \times 10^{11}$cm and $e \approx 0.14$. As these
parameters did not notably evolve, our analysis is insensitive to
the exact age of the pulsar.

Dewi and van den Heuvel \cite{ref3} and Willems and Kalogera
\cite{ref4} considered a scenario in which the progenitor star
lost most of its envelope through interaction with its companion
\A. Prior to the formation of the second pulsar, tidal interaction
between the progenitor and the neutron star has led to a circular
orbit. Since most of the lost mass could not have accreted onto
the companion, it was probably lost through a common envelope
phase, at which point the companion J0737-3039A was spun up and
its magnetic field was suppressed by accretion. They estimate that
the progenitor mass was more massive than $2.3~ \Ms$ \cite{ref3}
or $2.1~\Ms$ \cite{ref4} respectively. This limit follows from
standard evolutionary scenarios, leading neither to neutron star
formation nor to core collapse, from progenitors that are less
massive than $2.1-2.3~\Ms$ \cite{ref5}. The formation of the
second pulsar is described according to the current picture by a core
collapse event that involves a supernova and mass ejection from
the system.


We consider now the influence of mass ejection during the
formation of the second pulsar, on the orbital motion. The thrust
of the ejected mass, $\Delta m$, gives a velocity, $v_{cm}$, to
the center of mass (CM) of the remaining system. In addition, the
mass loss would lead to an elliptic orbit or even to the
disruption of the system. For a spherically symmetric mass loss,
$v_{cm}$ will be:
\begin{equation}
v_{cm} = \left( m_{Bi} \Delta m \over \left( m_A +
m_B\right)^{3/2} \left( m_A + m_{Bi} \right)^{1/2}\right) v_K
\label{vcm}
\end{equation}
where $v_K \equiv \sqrt{G (m_A + m_B) / R}$ is the {Keplerian
velocity of the two stars relative to each other, just after the
explosion, with $R$ being the distance between the two stars at
that time}. Within the context of J0737-3039, $m_A = 1.377(5)\Ms$,
$m_B = 1.250(5) \Ms$, $\Delta m$ is, of course, unknown and
$m_{Bi} \equiv m_B + \Delta m$ is the initial mass of \B, while
$v_K \approx 600$~km/s.  With $m_A \approx m_B $, $v_{cm}$ would be
of the order of $v_K/2$ unless $\Delta m \ll m_B$. The birth of
the system  with a low eccentricity and $\Delta m
\approx m_B$ requires for \B\ to have had a natal kick \cite{ref6}. This kick will
increase $v_{cm}$ further and the above value can be considered as
a lower limit. In other words, the nearly circular orbit today
implies either a large CM velocity, roughly as given by Eq.
\ref{vcm}, or a small ejected mass $\Delta m \ll m_B$.

\begin{figure}[tbh]
\center{ \epsfig{file=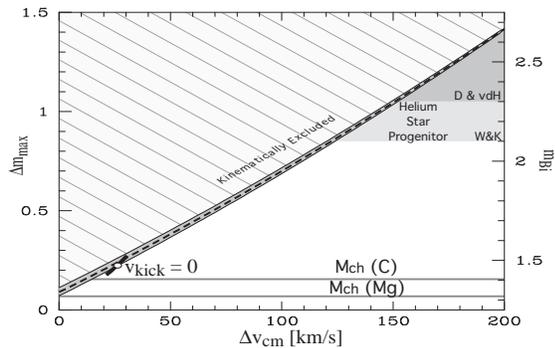,width=2.9in,height=1.8in}
 }
\vskip -3.8mm
\caption{ 
{The maximal kinematically allowed mass loss $\Delta m_{max}$ and
progenitor mass $m_{Bi}$ for a given  change in the CM velocity
$\Delta v_{cm}$. The upper mass depends on the actual eccentricity
($0.088< \ee < 0.14$), hence the finite width of the line. The
hatched area is kinematically excluded. The region
allowed by standard evolutionary scenario through the formation of
a He-star, which requires a minimum mass of $2.1~\Ms$ (marked W\&K
\cite{ref4}) to $2.3~\Ms$ (marked D\&vdH \cite{ref3}) to form a NS
through a core collapse SN, is marked as shaded triangles. The
no-kick solution that results with the above range of
eccentricities is marked by the short heavy line.}}

\end{figure}

Fig.\ 1 depicts change in the CM velocity (relative to
the unknown CM velocity prior to the explosion) for a system with two
masses moving on a circular orbit,  assuming the new binary system
attains the initial eccentricity of J0737-3039 ($0.088 < \ee <
0.14$).
In the following we assume that this initial CM velocity was small
and we approximate $v_{cm} \approx \Delta v_{cm}$. We will return
to this point in the conclusions. Fig.\ 1 shows that the
minimal CM velocity for a $2.1~\Ms$ progenitor is 120~km/s.


Recently Ransom et al.\ \cite{ref11new} have estimated
$v_{cm\perp}$, the CM velocity of the binary on the plane of the
sky, using the observed scintillations of the system. They find a
rather large  value: $v_{cm\perp}=141 \pm 8.5$~km/s (with $96.0
\pm 3.7$ km/s along the orbit and $103.1 \pm 7.7$~km/s
perpendicular to it). This value excludes the region in Fig.\ 1
left of a vertical line of $\sim$141~km/s. However, these findings
were questioned recently by Cole et al. \cite{ref12new} who
suggested that the scintillation pattern is anisotropic. When
including anisotropy, they find a much lower value:
$v_{cm\perp}=66 \pm 15$~km/s.  The region in Fig.\ 1 to the right
of the vertical line of $\sim$66~km/s is consistent with this
observation.

However, the system can be additionally constrained. The
observed distance of the system from the Galactic plane, $z_{obs}
\approx 50$~pc, enables us to place a statistical upper limit on
$v_{cm}$. Stars move in a periodic motion in the vertical
direction. For small vertical oscillations, the potential of the
Galaxy is harmonic: $\Phi=2\pi G\rho_0 z^2$, where $\rho_0 \approx
0.25~\Ms/$pc$^3$ is the mass density in the disk \cite{ref11}.
This gives a vertical orbital period, $P_z \approx 50$~Myr. The
typical velocity for an object at $z_{obs}$ is $v_z \approx 2\pi
z_{obs}/P_z$. $z_{obs} \approx 50$~pc implies then that the
expectation value of the vertical velocity is of the order of
6~km/s.

To quantify the probability for having a particular CM velocity
given the observed $z_{obs}$ we perform  Monte Carlo simulations
that follow the formation of the system. We assume  that star \B\
had a given mass $m_{Bi}$, and that a randomly oriented kick ${\bf
v}_{kick}$ was given to it. We also assume that the progenitor
distribution has an initial Gaussian distribution in the amplitude
of the vertical oscillation, with a width $\sigma_z = 50$~pc
(other $\sigma_z \lesssim 100$~pc gave very similar results). At
the moment of formation, we assume it had a random phase within
its vertical motion. We calculate the CM kick velocity ${\bf
v}_{cm}$, and assign it a random direction, then integrate the
vertical motion of the pulsar for $50$~Myr using a realistic
galactic potential \cite{ref12}.

Fig.\ 2a depicts the probability that a system with a given
$m_{Bi}$ and $v_{kick}$ could find itself with $0.087<  \ee< 0.14$
after $50$~Myr, within $50$~pc of the galactic plane and with a
transverse velocity of $66\pm 15$~km/s as measured by Coles et
al.\ \cite{ref12new}. For comparison, Fig.\ 2b depicts similar
simulations but without the assumption on the transverse velocity.
Here the initial conditions are less constrained. Fig.\ 4c repeats
the first panel, with the modified condition that the $\ee < 0.14$
instead of  $0.087< \ee < 0.14$. One can see that qualitatively
the results do not change.  Fig 4d is a repeat of the first panel
with the transverse velocity assumed to be $141 \pm 8.5$~km/s, as
measured by Ransom et al.~\cite{ref11new}, with the scintillation
anisotropy neglected.

\begin{figure}[t]
\vskip -0.1cm \center{ \epsfig{file=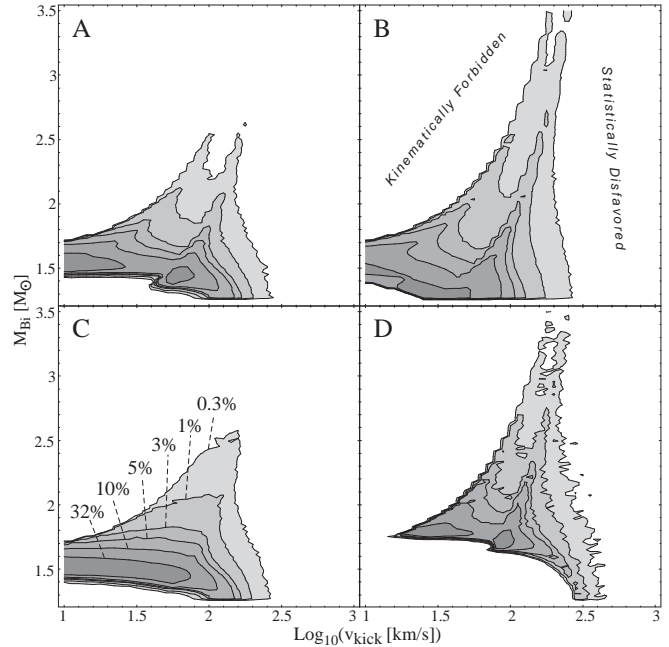,width=3.4in} \vskip
-0.25cm
 }
\caption{ 
{ Panel A: The probability that a binary system will end up within
$50$~pc from the galactic plane, with $0.088 <\ee < 0.14$ and with
a transverse velocity of $66 \pm 15$~km/s, given that $50$~Myrs
before, the progenitor system had a circular orbit, that star \B\
had a progenitor mass $m_{B,i}$, and it obtained a random kick
velocity of size $\vk$. The probabilities are normalized to the
most likely $M_{Bi}, \vk$. The He-star solution requires $m_{Bi}$
larger than about $2.1\Ms$. With a fine tuned $\vk$, this type of
a solution is ruled out at the 99\% c.l. Solutions with either
$m_{Bi} \approx 1.55 \pm 0.2~\Ms$ and $\vk \lesssim 30$~km/s, or,
$m_{Bi} \approx 1.45 \pm 0.2~\Ms$ and $ \vk\sim 50-80$~km/s, are
kinematically more favorable. In panel B,  the constraint on the
transverse velocity is alleviated. Here, He-star scenarios can be
ruled out only at the $97$\% c.l. The top left side is
kinematically forbidden (no solution without a large $\vk$ exists
to counter the ejected mass), while the r.h.s.\ is statistically
disfavored since a large $\vk$ implies a low probability for
finding the system near the galactic plane. Unlike Panel A, the
eccentricity in panel C is constrained to be $\ee <
0.14$---showing that double ridged solution arises from the
constraint on eccentricity. Panel D is the same as panel A, except
that here the transverse velocity is $141 \pm 8.5$~km/s. Note that
as marked on Panel B while the region on the upper left side -
high mass and low kick velocity is kinematically forbidden, the
region on the right hand side and in particular the top right hand
side - high mass and large kick velocity is allowed kinematically
it is just disfavored statistically. }}
\end{figure}

Without the additional information on the  CM velocity, we find
that that even if $v_\mathrm{kick}$ is fine tuned to be near
either 130 or 315~km/s, a system with a mass of $m_{Bi}=2.1~\Ms$
could result with the observed configuration in only about 3\% of
the random realizations (as compared with the most favorable
conditions having lower $m_{Bi}$ and $v_\mathrm{kick}$).  Other
$\vk$'s, or higher mass systems are kinematically even less
likely. On the other hand, low ejected mass solutions are favored.
If we add the constraint that the transverse velocity is $66 \pm
15$~km/s, then a fine tuned $m_{Bi}=2.1~\Ms$ model can be ruled
out at even 99\% c.l. The results, are though, qualitatively
different if we take the CM velocity on the plane of the sky as
$141$~km/s. Much larger kicks, of a few hundred km/s are essential
now. While the ``cannonical" $2.1\Ms$ or higher  solution is not
really favorable yet, the previously most favorable  low mass
($1.4 \Ms$) progenitor is ruled out. Instead an ``intermediate"
mass of $\sim 1.7 \Ms$ is the most likely one.

If we assume that the transverse measurement of Coles et al.\
\cite{ref12new}, which allows for the scintillation to be
anisotropic, then at better than 99\% confidence, the progenitor
of star \B\ was less massive than $2.1~\Ms$. This is just the
lower limit of Willems and Kalogera \cite{ref4} and it is slightly
below the lower limit of Dewi and van den Heuvel \cite{ref3} for
the progenitor mass in a standard core collapse scenario. An
inspection of Fig.~2a reveals that there is a kinematically
favorable solution with a small mass loss and natal kicks ranging
from 0 to $\sim 100$~km/s. Since a small mass loss necessarily
implies a small natal kick, the solutions with $v_{kick} \gtrsim
100$~km/s are physically unlikely. If we therefore limit ourselves
to $v_{kick}\lesssim 30$~km/s, then without fine tuning, a large
fraction of the progenitor systems will result with the observed
configuration with a progenitor mass of $1.45~\Ms \lesssim m_{Bi}
\lesssim 1.65~\Ms$. Thus, a mass loss of about $0.3 \pm 0.1~\Ms$
is most probable. The results are qualitatively the same even if
we do not enforce that the resulting system has a CM velocity of
66~km/s, or if we replace the condition on the eccentricity to
$\ee < 0.14$ rather than $ 0.088 <\ee <0.14$.

At a distance of 600~pc a  velocity on the plane of the sky,
$v_{cm\perp}$, implies a proper motion of
$0.036(v_{cm\perp}/100~$km/s$)\phantom{.}\arcsec/\mathrm{yr}$.
This should be compared with the current errors of $\sim
0.04\arcsec$\ in the position of the system. This comparison
suggests that within a year or so we could obtain a direct limit
on the peculiar motion. A comparison of this peculiar motion with
the motion measured using scintillations would confirm the
anisotropy estimates that arise from the scintillations
measurements.  The break down to the different components should
allow us to measure the orientation of the orbital plane in the
sky. Thus within a year or two we should be able to nail down the
two components of the CM motion on the plane of the sky as well as
the orientation of the orbital plane!  The knowledge of the
vertical component (relative to the galactic plane) would tell us
whether the position of the  system in the galactic plane is
natural (if the vertical component is small) or an unlikely
coincidence (if this velocity is large).

We are left with two physically distinguished scenarios for the
formation of pulsar \B. In the first, the progenitor is a
kinematically ``unlikely" but theoretically plausible $2.1-
2.3~\Ms$ He-star progenitor---around the minimal masses estimated
from stellar evolution scenario \cite{ref3,ref4}. In the second
scenario, the progenitor is a kinematically favorable $\sim
1.5~\Ms$ young stellar core. The more probable low mass solution
requires a new type of stellar collapse. Intermediate solutions are
neither statistically favored nor do they fit any plausible
theoretical scenario.

Before turning to the implications of these two solutions we
consider, first, three assumptions made in our analysis. (i) We
have assumed that prior to the formation of the second pulsar the
system was in a circular motion. This  follows from all
evolutionary scenarios that lead to a neutron star and a small
mass progenitor (even a $2.3 \Ms$ is a very small mass
progenitor). (ii)~We have assumed that the second mass loss was
instantaneous, namely, shorter  than a fraction of an orbital
period. Given that the orbital motion is of several hundred km/s
while typical mass ejection velocities in SNe are higher than
10,000~km/s, this assumption is reasonable (for all conventional
neutron star formation scenarios). (iii) We have assumed that the
CM velocity, prior to the formation of the second pulsar was
small. One would expect that the system would have acquired a CM
velocity during the formation of the first pulsar. About half of
the system's mass was lost during this event and this should have
resulted in some CM velocity. However, this velocity would have
been of the order of half the Keplerian velocity of the system at
that time (see Eq. \ref{vcm}). Given the fact that the orbital
separation was much larger we could reasonably expect, but not
prove, that this velocity was of order of several tens of km/s.
Moreover, if the system would have acquired a large CM velocity in
the first SN, there would have been an even smaller probability to
find it in the galactic plane today.

We turn now to the most likely scenario, the very low mass
scenario. We imagine the same evolutionary scenario in which some
time before 50 Myr, system \A\ and progenitor \B\ were in a common
envelope phase and \B\ lost most of its mass keeping practically
just its core of $\sim 1.45~\Ms$. This progenitor leads to a small
CM motion and does not require any kick velocity. This solution is
kinematical preferred. However, it requires a new mechanism for
the formation of the pulsar as He-stars of $1.45~\Ms$ do not
collapse to form neutron stars \cite{ref5,ref13}.  The observation
that the progenitor mass is very close to the Chandrasekhar mass
leads us to conjecture that the process involves the collapse of a
supercritical white dwarf. For example, the progenitor may have
been a degenerate bare core just above the critical Chandrasekhar
mass, supported by the extra thermal pressure against collapse. As
it cooled, the additional support was lost and the core collapsed
to form a neutron star. A second possibility is that it was formed
just below the Chandrasekhar mass, and as it cooled,
neutronization at the core increased the baryon to electron ratio,
and with it reduced the Chandrasekhar mass until the progenitor
became unstable and collapsed. Note that the object must have been
composed of O-Ne-Mg as a collapsing CO core would have
carbon-detonated and it would have exploded completely forming a
type I SNe and leaving no remnant. This solution clearly requires
a new type of formation scenario for neutron stars.

This new solution passes an immediate non trivial test. With a
small mass loss, only small kick velocities are possible and in
fact we can estimate in this case (see Fig.\ 2) the mass loss
needed to obtain the initial eccentricity $e \approx 0.11$. We
find (while conservatively assuming that the collapse took place
$50^{+100}_{-50}$~Myr ago) $\Delta m = e(1+q) m_B = 0.28 \pm
0.07~\Ms$  and a progenitor mass of $1.53\pm 0.07~ \Ms$, which is,
indeed, just above the Chandrasekhar limit. Intriguingly, some
mass loss, in the form of $\nu$ losses of a few times
$10^{53}$ergs, must take place. The estimated $\Delta m$
corresponds to $E_\nu \approx \Delta mc^2 \approx 4.2 \times
10^{53}$ergs, which is in the right range. Of course, if some mass
is ejected as well, $E_\nu$ will be smaller and a small kick could
arise, but the total mass-energy lost will be the same. The
consistency of this mass and energy loss with the previous
physical picture increases our belief in this new and unusual
scenario.

The other, higher mass scenario of $2.1-2.3\Ms$  (which is still
on the lowest end of the evolutions scenarios) is the most likely
from a stellar evolution point of view. However, as we have seen
it is quite unlikely statistically. Specifically, it implies a
post-kick CM velocity of at least 125 to 150 km/s, which was
within a few degrees from the Galactic plane for it to find itself
in the plane today. To be consistent with the low $v_{cm\perp}$,
it should also be pointing within about 25$^\circ$ from us, or in
the opposite direction. In other words, it should have a minimum
velocity of at least 100 km/s in the direction of the line of
sight, if progenitor \B\ was a 2.1$\Ms$ star. More massive
progenitors imply even larger velocities. Of course it cannot be
ruled out. We might be observing a very improbable system.
However, if this is the case, the ``unlikeliness" of the system
should be taken into consideration when performing population
synthesis and estimating neutron star mergers rate based on this
observation. Specifically, with a very large CM velocity, as such
a system must have, it is very unlikely to find it in the Galactic
plane. Most such systems should exist high above the Galactic
disk. Anticipating this, Narayan, Piran \& Shemi \cite{ref13new}
considered this effect by taking an effective scale hight of 5kpc
for the Galactic binary NS systems. This value was used later by
others. It may be that even this high value is too low. This in
turn might boost up the estimated rates of NS merger by a
significant factor.

The authors thank Ramesh Narayan, Ehud Nakar, Re'em Sari  and
members of the Israeli center for High Energy Astrophysics, D.
Eichler, M. Milgrom, E. Waxman and V. Usov for helpful comments.
We thank Steve Thorsett for pointing out an error in an earlier
version of this manuscript.  The research was supported by an ISF
grant.

\def\mnras{Mon.\ Not.\ Roy.\ Astr.\ Soc.}
\def\apj{Ap.\ J.}
\def\apjl{Ap.\ J.}
















\end{document}